\documentclass[MSNbibl,dvips]{arxstspdf}
\usepackage{multirow}
\usepackage{graphicx}

\usepackage{flushend}
\usepackage{stfloats}

\volume{26}
\issue{3}
\pubyear{2011}
\firstpage{388}
\lastpage{402}
\doi{10.1214/11-STS361}



\begin{document}
\begin{frontmatter}

\title{Misspecifying the Shape of a Random Effects Distribution: Why Getting It Wrong May Not Matter}
\runtitle{Misspecifying a random effects distribution}

\begin{aug}
\author{\fnms{Charles E.} \snm{McCulloch}\corref{}\ead[label=e1,text={chuck@biostat. ucsf.edu}]{chuck@biostat.ucsf.edu}}
\and
\author{\fnms{John M.} \snm{Neuhaus}\ead[label=e2,text={john@biostat. ucsf.edu}]{john@biostat.ucsf.edu}}
\runauthor{C. E. McCulloch and J. M. Neuhaus}

\affiliation{University of California, San Francisco}

\address{Charles E. McCulloch is Professor and Head, Department of
Epidemiology and Biostatistics,
185 Berry Street, San Francisco, California 94107, USA \printead{e1}.
John M. Neuhaus is Professor,
Department of Epidemiology and Biostatistics,
185 Berry Street, San Francisco, California 94107, USA \printead{e2}.}

\end{aug}

%
\begin{abstract}
Statistical models that include random effects are commonly used to
analyze longitudinal and correlated data, often with strong and
parametric assumptions about the random effects distribution. There is
marked disagreement in the literature as to whether such parametric
assumptions are important or innocuous. In the context of generalized
linear mixed models used to analyze clustered or longitudinal data, we
examine the impact of random effects distribution misspecification on a
variety of inferences, including prediction, inference about covariate
effects, prediction of random effects and estimation of random effects
variances. We describe examples, theoretical calculations and
simulations to elucidate situations in which the specification is and
is not important. A key conclusion is the large degree of robustness
of maximum likelihood for a wide variety of commonly encountered
situations.
\end{abstract}

%

\vspace*{-3pt}
\end{frontmatter}

\section{Introduction}
\vspace*{-3pt}
Statistical models that include random effects are commonly used to
analyze longitudinal and clustered data. When generalized linear mixed
models (McCulloch, Searle and Neuhaus, \citeyear{mccusearneuh2008}) and maximum likelihood estimation are
used to analyze such data, strong, parametric assumptions about the
random effects distribution are typically made. Are inferences
sensitive to this specification?

One body of research has indicated that the impact of\vadjust{\goodbreak}
misspecification of the \textit{shape} of the random effects distribution is slight,
especially for estimating regression parameters other than the
intercept
(e.g., Neuhaus, Hauck and Kalbfleisch, \citeyear{neuhhauckalb1992}; Neuhaus, Kalbfleisch and Hauck,
\citeyear{neuhkalbhauc1994};
Butler and Louis, \citeyear{butlloui1992}; Heagerty and Kurland, \citeyear{heagkurl2001}).

However, a number of authors have claimed sensitivity to parametric
specification of a random effects distribution. An oft-quoted article
is that of Heckman and Singer (\citeyear{hecksing1984}), which (in describing performance of
maximum likelihood estimation) states, ``Both theoretical and empirical
examples indicate that estimates of structural parameters obtained\break from
conventional procedures are very sensitive to the choice of mixing
distributions.'' Other authors have suggested more flexible
distributional assumptions for the random effects to protect against
misspecification. Approaches include nonparametric ma\-ximum likelihood
(Aitkin, \citeyear{aitk1999}; Agresti, Caffo and Ohman-Strickland, \citeyear{agrecaffohma2004}), more
flexible parametric
distributions (Zhang et~al., \citeyear{zhansongqugree2008}), marginalized mi\-xed effects
models (Heagerty and Zeger, \citeyear{heagzege}), h-li\-kelihood approaches that can be easily
adapted to fit different distributions (Lee and Nelder, \citeyear{leeneld2004}), families of
parametric distributions (Piepho and McCulloch, \citeyear{piepmccu2004}), mixtures of normal
distributions (Lesaffre and Molenberghs, \citeyear{lesamole1991}) and ``smooth'' nonparametric fits
(Zhang and Davidian, \citeyear{zhandavi2001}). A recent article (Huang, \citeyear{huan2009}) encapsulates the
sentiment used to justify these approaches:

\begin{quote}
For computational convenience, random effects in GLMMs are almost
routinely assumed to be normal. However, the normality assumption may
be unrealistic in some applications. \ldots Early investigation to
address this concern suggested that misspecifying the models for the
random effects usually only results in a small amount of bias in the
maximum likelihood estimators (MLEs) for the fixed effects
(Neuhaus, Hauck and Kalbfleisch, \citeyear{neuhhauckalb1992}). However, more recently, many authors have
found that likelihood-based inference can be se\-verely affected if the
random-effect model is misspecified. For example,
Heagerty and Kur\-land (\citeyear{heagkurl2001}) computed the asymptotic bias in the MLEs for
the parameters in~a~lo\-gistic mixed model in four instances
of~ran\-dom-effect model misspecification. They concluded that incorrect
assumptions on the random effects can lead to \mbox{substantial} bias in the
MLEs for the fixed effects. Ag\-resti, Caffo and Ohman-Strickland (\citeyear{agrecaffohma2004}) conducted
empirical studies on the impact of model misspecification for the
random effects in GLMMs, showing that the MLEs for the fixed effects
can be very sensitive to the assumed random effect model. Finally,
Liti{\`{e}}re, Alonso and Molenberghs Li\-ti{\`{e}}re, Alonso and Molenberghs (\citeyear{litisaskmole2007})
used simulation to show that the type I
and type II errors of tests for the mean structure in a logistic mixed
model can be seriously affected by violations of the random-effect
model.
\end{quote}

A body of work with a slightly different focus has been that of
estimating the shape of the random effects distribution, also by
hypothesizing more flexible distributional fits for the random effects.
Though this is different from assessing misspecification, it is a
closely related problem and those methods may help diagnose
misspecification in the situations in which sensitivity to
misspecification is of concern. Approaches to this problem have used
mixtures of normal distributions (Magder and Zeger, \citeyear{magdzege1996}; Caffo, An and Rohde,
\citeyear{caffabrhod2007}),
and nonparametric and ``smooth'' nonparametric fits
(Laird, \citeyear{laird1978}; Davidian and Gallant, \citeyear{davigall1993}; Zhang and Davidian,
\citeyear{zhandavi2001}; Ghidey, Lesaffre and Eilers, \citeyear{ghidlesaeile2004}). See
Ghidey, Lesaffre and Verbeke (\citeyear{ghidlesaverb2010}) for a recent review article in the context of
linear mixed models.

We use data from the Heart and Estrogen Replacement Study (HERS)
(Hulley et~al., \citeyear{hulleyhers1998}) to illustrate some of our findings. HERS was a
randomized, blinded, placebo controlled trial for women with previous
coronary disease. The study enrolled 2,763 women and followed them
annually for five subsequent visits, generating longitudinal data. More
detail is given in Section \ref{sechers}, but we foreshadow the results
here. We fit logistic regression models with random intercepts to
model whether or not a woman had high blood pressure as a function of
her body mass index, whether she was on hypertensive medication, and a
trend over the visits. For the random intercepts the models assumed
one of four distributions: normal, a centered and scaled exponential
distribution, a $\operatorname{Tukey}(g,h)$ distribution or a 3 point discrete
distribution, quite different parametric assumptions. The regression
parameters were very similar, despite evidence for differences in
overall model fit. How can we reconcile this with the claims above?

We will argue that concerns over the parametric specification of random
effects distributions are sometimes valid, but ofttimes misplaced. This
is due to three reasons: (1) sensitivities are restricted to aspects
of the estimation that are not typically of interest, (2) the
situations considered are unfairly extreme, or (3) that close scrutiny
of published results does not actually support sensitivity to
misspecification.

Our paper is organized as follows. In the next section we outline a
series of examples and their associated inferences. We then describe
the basic statistical model. The sections that follow look at the
impact of misspecification on various aspects of the model. The
remaining sections consider some simulation studies, the HERS example
(in more detail), some remarks on more complicated random effects
structures, and we conclude with a brief summary.

\section{Inferential Settings}

A key argument we make is that the robustness is dependent on the
inferential setting. Perhaps the most common inference for which a
regression model is used is estimation and testing of regression
coef\-ficients associated with a covariate. In the clustered data
setting, a useful distinction is whether a covariate is a~``between-cluster''
covariate, meaning that it is constant over the
units in a cluster or a ``within-cluster'' covariate, meaning that it
varies within\break a~cluster, but has an average that is constant between
clusters. For example, in a longitudinal study with participants being
clusters and all participants contributing data across 4 time points,
the ethnicity of the subject would be a between-cluster covariate and
the visit number would be a~within-cluster covariate. Of course,
covariates are often neither purely between or within, for example, the
body mass index of a participant measured over time, in which case it
is often useful for pedagological, substantive or technical reasons to
decompose a~covariate into its purely-between and purely-within
components (Neuhaus and Kalbfleisch, \citeyear{neuhkalb1998}; Raudenbush and
Bryk, \citeyear{raudbryk2002}; Neuhaus and McCulloch, \citeyear{neuhmccu2006}). Other
inferential goals include using the fitted model to generate
predictions, generating predicted values of the random components of
the model, or estimating aspects of the distribution of the random
components of the model, for example, as in a variance components
analysis (Searle, Casella and McCulloch, \citeyear{searcasemccu}).

\subsection{Examples}
To illustrate these distinctions, we begin with a~series of examples.

\subsubsection{Estimate the effect of a within-cluster covariate}

Metlay et~al. (\citeyear{metlcamamackmccu2007}) evaluated the effectiveness of an
educational program that attempted to reduce the inappropriate use of
antibiotics for conditions nonresponsive to antibiotics. The study was
a cluster randomized trial, randomized at the level of the hospital,
with 8 intervention and 8 control hospitals. Measurements were taken
at each hospital in the year before and after the interventions were
introduced. The data were further clustered by physician within
hospital. The primary inference involved a comparison of responses
before and after the intervention, a within-cluster ($={}$hospital)
covariate setting.

\subsubsection{Estimate the effect of a between-cluster covariate}

In the previous example, 8 of the hospitals were Veteran's
Administration (VA) hospitals and 8 were non-VA. A secondary goal was
to compare the rates of inappropriate usage between VA and non-VA
hospitals, a between-cluster covariate setting.

\subsubsection{Derive predictions based on the fixed effects}

Auble et~al. (\citeyear{aublhsiemccayeal2007}) studied the performance of four clinical
prediction rules for estimating the risk of death or serious
complications for patients admitted to the hospital with a diagnosis of
heart failure. The goal was to develop a rule that could stratify
patients into low and high risk groups. One of the models used was a~random
effects logistic regression model accommodating clustering of
outcomes by hospital. The inferential goal was to develop a prediction
rule based on the fixed effects in the model.

\subsubsection{Predict random effects} Zhang et~al. (\citeyear{zhansongqugree2008})
wanted to estimate subject-specific rates of disease progression in
chronic kidney disease patients. They estimated subject-specific slopes
of the change in glomerular filtration rate (a measure of kidney
function) over time. They used a linear mixed model with clustering by
subject and non-normally distributed random effects. The inferential
goal was to predict the realized values of the random effects.

\subsubsection{Estimate variance components}

Selby et~al. (\citeyear{selb1996}) studied the variation in the rates of coronary
angiography (a diagnostic procedure using X-rays and a~special dye to
diagnose coronary artery disease) across 16 hospitals. Significant
variation after adjusting for differences in patient populations
indicates over- or under-use of the procedure, leading to poor patient
outcomes and/or wasted resources. The inferential goal was to estimate
the variance of the hospital random effects after adjusting for patient
characteristics.

\subsection{Longitudinal and Clustered Studies}

Each of the above examples describes clustered or longitudinal studies,
in which repeated observations of the outcome are taken within
clusters, for example, multiple measurements of kidney function on
a~patient. These are the types of scenarios we consider.

An immediate distinction to make is the different scenario considered
by Heckman and Singer (\citeyear{hecksing1984}), in which there is only a single observation of
the outcome per cluster (in their case, a time-to-event outcome). In
the normal linear mixed model setting, with only a single observation
per cluster, the random effect and error term are completely
confounded, leading to lack of identifiability. The presence of
identifiability in the Heckman and Singer situation arises
only through strong parametric assumptions, so it is not surprising\vadjust{\goodbreak}
that results are sensitive to the choice of distribution. Such
settings do not represent longitudinal or clustered studies and we do
not consider them further.

\section{A Generalized Linear Mixed Model}

The model we will use to assess the impact of misspecification is a
generalized linear mixed model for clustered data with random
intercepts, $b_i$. There are fewer results for more complicated
covariance structures, such as those generated by random intercept and
slope models, which we discuss briefly in Section \ref{secrslopes}. Let
$Y_{it}$ represent the $t$th observation $(t=1, \ldots, n_i)$ within
cluster $i$ $(i=1, \ldots, m)$. We assume that, conditional on the
random effects, the~$Y_{it}$ are independent:
%
\begin{eqnarray}\label{glmm}
Y_{it}|b_i &\sim&\mbox{ independent } F_Y,\nonumber\\
  \eqntext{i=1, \ldots, m; t=1, \ldots, n_i,}\\
g (\mbox{E}[Y_{it}|b_i] )&=& b_i + \mathbf{x}'_{it}\bolds
{\beta},\\
b_i &\sim& \mbox{ i.i.d. } F_b , \nonumber\\
\mbox{E}[b_i] &=& 0 \quad\mbox{and}\quad \operatorname{var}(b_i) = \sigma
^2_b\nonumber,
\end{eqnarray}
where $g(\cdot)$ is a known link function, $\bolds{\beta}$ is the
parameter vector for the fixed effects, and $\mathbf{x}_{it}$ is a vector
of covariates for cluster $i$ at time $t$. We consider the performance
of maximum likelihood for estimating the parameters. We focus
especially on the situation where the assumed distribution is normal,
since this is the most commonly implemented choice in popular software
packages such as SAS, Stata and R.

The model, (\ref{glmm}), contains a number of specifications. The main
specification we will consider is the \textit{shape} of the random
effects distribution, that is, the parametric form of the distribution,
$F_b,$ may be incorrectly specified.

We have briefer comments on two other aspects of the specification. We
may be concerned that:
\begin{itemize}
\item Basic characteristics of the random effects distribution may
depend on a covariate. For example, the mean or variance of $F_b$
depends on a covariate.
\item There is a dependence of the random
effects distribution on cluster sample size. For example, the mean of
$F_b$ depends on $n_i$.
\end{itemize}

When the mean of the random effects distribution depends on a
covariate, a fundamental relationship is introduced between the
covariate and the distribution, potentially creating a serious bias in
estimating the form of the relationship between the covariate and the
outcome. Neuhaus and McCulloch (\citeyear{neuhmccu2006}) discuss reasons for this, suggest
decomposing covariates of interest into within- and between-cluster
components and investigate conditions under which this decomposition
can remove or reduce bias. Heagerty and Kurland (\citeyear{heagkurl2001}) study situations where
the variance of the random effects distribution depends on a covariate
and consider use of conditionally specified models (as we do here), as
well as models that specify a marginal regression equation (e.g.,
logistic regression for binary outcomes) but incorporate random effects
in an underlying conditional model. They show that, for estimating the
conditionally specified parameters of (\ref{glmm}), the impact of
highly unequal variances can lead to substantial bias.
So both of these aspects of the specification of~(\ref{glmm}) are
potentially important.

Some authors have argued (Hoffman, Sen and\break Weinberg, \citeyear{hoffsenwein2001}; Williamson,
Datta and Satten, \citeyear{willdattsatt2003}) that
when the cluster sample size is incorrectly assumed to be independent
of the random effects distribution, serious consequences may result.
Neuhaus and McCulloch (\citeyear{neuhmccu2011}) argue that this type of misspecification is really
a variation of incorrect assumptions about the mixing distribution. If
one assumes that the parameters in the joint distribution of $n_i$ and
$X_{ij}$ are not functionally related to the parameters of interest,
namely, those in the distribution of outcomes or random effects, then
as a~function of $\beta$, Neuhaus and McCulloch (\citeyear{neuhmccu2011}) show that
%
\begin{eqnarray}\label{eqtruecondit1}
&&f(\mathbf{y}_{i}, n_i , \mathbf{X}_{i})\nonumber\\
 &&\quad \propto
\int_b \Biggl\{ \prod_{j=1}^{n_i} f_Y(y_{ij} | n_i, X_{ij}, b) \Biggr\}\,dF_{b}(b | n_i, \mathbf{X}_{i} )\\
&&\quad =
f(\mathbf{y}_{i} | n_i , \mathbf{X}_{i}). \nonumber
\end{eqnarray}

This is a useful representation of the joint likelihood of $\mathbf{y}_{i}, n_i$ and
$\mathbf{X}_{i}$ since analysts would typically model
the distribution of the responses conditional on covariates and sample
size. In particular, ignoring the association of cluster size with
response, one would base inference on the incorrect likelihood built up
of terms
%
\begin{equation}\label{eqfalseglmm}
f^*(\mathbf{y}_{i} | \mathbf{X}_{i}, n_i) =\int_b \prod_{j=1}^{n_i} f_Y
(y_{ij} | X_{ij}, b)\,dF^*_b(b),
\end{equation}
where the asterisks denote assumed distributions.

Comparing equation (\ref{eqfalseglmm}), the assumed likelihood and
(\ref{eqtruecondit1}), the true likelihood, we see that we can view~(\ref{eqfalseglmm})
as arising from (\ref{eqtruecondit1}) but with a
misspecified random effects distribution, namely, the conditional
distribution of $b$ given $n_i$ and $\mathbf{X}_i$ is incorrectly
specified as $f^*_b(b)$. This is important because some of the comments
we make below for random intercepts models concerning misspecification
of the \textit{shape} of the random effects distribution will also
apply to incorrectly assuming that the cluster sample size is
independent of the random intercept distribution. Situations in which
the informative cluster size is related to the random slopes are more
involved.

We next consider in more detail the impact of shape misspecification of
the random effects distribution, organized by the inferential targets.

\section{Estimate/Test a Within-Cluster Covariate}

Virtually every study of the impacts of misspeci\-fication has shown
little impact on within-cluster~co\-variates. Neuhaus, Hauck and Kalbfleisch (\citeyear{neuhhauckalb1992}),
Neuhaus, Kalbfleisch and Hauck (\citeyear{neuhkalbhauc1994}) and Heager\-ty and Kurland (\citeyear{heagkurl2001})
developed analytic
results to assess the impact of misspecification of the random effects
distribution using the theory of inference under misspecified models
(Huber, \citeyear{hube1967}; Akaike, \citeyear{akai1973}; White, \citeyear{whit1994}). This theory shows that estimators
obtained by maximizing the likelihood based on a misspecified random
effects distribution converge to values that minimize the
Kullback--Leibler divergence (Kullback, \citeyear{kull1959}) between the correctly
specified and misspecified models. In cases such as the linear mixed
effects model and binary matched pairs (Neuhaus, Kalbfleisch and Hauck, \citeyear{neuhkalbhauc1994}), one can
obtain closed form solutions for the values that minimize the
Kull\-back--Leibler divergence. In these cases, the closed form solution
shows that the estimator based on~mis\-specified random intercepts is
identical to the standard conditional likelihood estimator, an
estimator unaffected by the random effects distribution. In other
cases, one can show consistent estimation at $\beta=0$ for generalized
linear mixed models with misspecified random intercept distributions.
For example, Neuhaus, Hauck and Kalbfleisch (\citeyear{neuhhauckalb1992}) show consistent estimation at
$\beta=0$ for logistic models with misspecified random intercept
distributions and Neuhaus and McCulloch (\citeyear{neuhmccu2011}) extend the result to the entire
class of generalized linear mixed models. In addition, one can obtain
approximate solutions to Kullback--Leibler divergence minimizers using
Taylor expansions (Neuhaus, Hauck and Kalbfleisch,\break \citeyear{neuhhauckalb1992}). Using this device,
Neuhaus, Hauck and Kalb\-fleisch (\citeyear{neuhhauckalb1992}) show little asymptotic bias in a logistic model
with misspecified random intercepts. Heagerty and Kurland (\citeyear{heagkurl2001}),\vadjust{\goodbreak} in their
Table 1, show~vir\-tually no impact on the asymptotic bias in logistic
regression models with a wide variety of gamma-distributed random
effects. Chen, Zhang and Davidian (\citeyear{chenzhandavi2002}), in investigating bias and efficiency
in logistic models, state that:

\begin{quote}
Estimation of $\beta_2$, which corresponds to a~covariate changing
within individuals, suffers no loss of efficiency under
misspecification of the random effects distribution. Similar results
have been reported by Tao et~al. (\citeyear{taopaltyandnewt1999}) and Zhang and Davidian
(\citeyear{zhandavi2001}). \ldots a within individual covariate such as time is roughly
`orthogonal' to among-individual effects. Thus, estimation of the
associated regression coefficient may be less affected.
\end{quote}
The above cited Zhang and Davidian article, investigating a linear
mixed model, shows little impact on bias or efficiency.

So for the inferential goal which is arguably the most relevant to
clustered or longitudinal data, misspecification of the shape of the
random effects distribution has little or no effect.

\section{Estimate/Test a Between-Cluster Covariate}
Between-cluster covariates might be expected to be more sensitive to
shape specification. We agree with Chen, Zhang and Davidian (\citeyear{chenzhandavi2002}), who state,
``We conjecture that, because a cluster-level covariate such as
treatment and the latent random effects both pertain to
among-individual variation, misspecification of the random effects
distribution would compromise quality of estimation of the
corresponding regression coefficient.'' However, the results of
Neuhaus, Hauck and Kalbfleisch (\citeyear{neuhhauckalb1992}) and Neuhaus and McCulloch (\citeyear{neuhmccu2011})
on consistent
estimation at $\beta=0$ and the Taylor approximations of
Neuhaus, Hauck and Kalbfleisch (\citeyear{neuhhauckalb1992}) apply equally to within- and between-cluster
covariates. In addition, the simulation results of Heagerty and Kurland (\citeyear{heagkurl2001})
show virtually no bias in estimates of between-cluster covariate
effects.

Via simulation, Agresti, Caffo and
  Ohman-Strick\-land (\citeyear{agrecaffohma2004}), in their Table 3 and for a
logistic model, showed moderate loss in efficiency when comparing
assumed normal and two-point discrete distributions versus the same two
true distributions with a small number of clusters (10). For a linear
mixed model, Magder and Zeger (\citeyear{magdzege1996}), in their Table 3, showed little
or no loss in efficiency when\vadjust{\goodbreak} assuming a normal random effects
distribution when the distribution was actually skewed or bimodal, but
moderate efficiency loss when the true distribution was a two-point
discrete distribution.

$\!\!\!$Of more concern are the results of Zhang and Davi\-dian (\citeyear{zhandavi2001}),
Liti{\`{e}}re, Alonso and Molenberghs (\citeyear{litisaskmole2007}) and Liti{\`{e}}re, Alonso and
Molenberghs (\citeyear{litialonmole2008}), who investigated
performance under highly non-nor\-mal distributions, but not as extreme
as a two-point discrete distribution. Zhang and Davidian (\citeyear{zhandavi2001}) present
simulation results for a linear mixed model with a true mixture of
normals distribution but fit with an assumed normal distribution and
show loss of efficiency. Liti{\`{e}}re, Alonso and Molenberghs (\citeyear{litisaskmole2007}) present simulation
results for a logistic mixed model under a~true ``Power function''
distribution (a continuous, but bounded range distribution) and assumed
normal distribution and claimed a significant loss of power under an
assumed normal distribution.

$\!\!\!$The scenario simulated by Zhang and Davidian~was a balanced, linear
mixed model, comparing clustered responses between two levels of a
binary between-cluster covariate. The usual maximum likelihood ba\-sed
analysis estimates the between group difference with the ordinary least
squares estimate (Searle, Ca\-sella and McCulloch, \citeyear{searcasemccu}), the difference in the group
means. So the comparison simulated was essentially the performance of
the t-test under non-normality. While the two-group t-test is well
known to be highly robust (e.g., Rasch and Guiard, \citeyear{rascguia2004}), it is not
impervious to non-normality and demonstrates loss of efficiency in this
extreme violation of the normality assumption. However, the estimates
are exactly unbiased.

$\!\!\!$The Liti{\`{e}}re, Alonso and Molenberghs (\citeyear{litisaskmole2007}) and~Li\-ti{\`{e}}re,
Alonso and Molenberghs (\citeyear{litialonmole2008}) simulations~in\-vestigate
the performance of tests that assume a~normal distribution
for the random effects while varying the true distributions. This is
useful for understanding the robustness of methods available in
standard software (which often assume normally distributed random
effects) when the true distribution is different than the assumed.

$\!\!\!$However, this does not directly address the question of
misspecification, which requires fixing the true distribution and
varying the assumed distribution. To see why this is the case,
consider a hypothetical situation where the standard deviation of an
estimated parameter is 1 when the true and assumed random effects
distributions are normal, but is equal to 2 when the true and\vadjust{\goodbreak} assumed
random effects distributions are $t$ with 3 degrees of freedom.
Further, suppose that, when the true distribution is $t_3,$ but we
assume the distribution is normal, the standard deviation is 2.1. In
this case, we would conclude that misspecification (assuming the
distribution is normal when, in fact, it is $t_3$) has little impact,
decreasing efficiency by $5\%$. But, if we compare the standard
deviations under the two true distributions (1~under normality and 2.1
under $t_3$), we might be tempted to incorrectly conclude that
misspecification causes a~large decrease in efficiency. In
Neuhaus, McCulloch and Boylan (\citeyear{neuhmccuboyl2010}), we therefore considered the same scenario as
investigated in Liti{\`{e}}re, Alonso and Molenberghs (\citeyear{litisaskmole2007}), but we varied the assumed
distribution. Contrary to their conclusions, our results showed that
misspecification had virtually no impact on power, except a very modest
impact for the combination of small sample sizes with huge random
effects variances (variances of 16 or 32 on the logit scale).
Liti{\`{e}}re, Alonso and Molenberghs (\citeyear{litialonmole2008}) consider the situation of random intercepts and
slopes, which we consider in Section \ref{secrslopes}.

Similar results are true of investigations of informative cluster sizes
with random intercepts. Hoffman, Sen and Weinberg (\citeyear{hoffsenwein2001}),
Williamson, Datta and Satten (\citeyear{willdattsatt2003}) and Benhin, Rao and Scott (\citeyear{benhrao2005})
investigate the impact
of informative cluster sizes on between-cluster covariates in logistic
models via simulation. Hoffman, Sen and Weinberg (\citeyear{hoffsenwein2001}) states that, ``We will
show by a simulated scenario that when cluster size is nonignorable the
behaviour of the generalised estimating equations approach breaks
down.'' and Williamson, Datta and Satten (\citeyear{willdattsatt2003}) claim that, ``\ldots the usual
generalized estimating\break equation approach resulted in severely biased
estimates of both the marginal regression and association parameters.''
These suggest a general failure of generalized estimating equation
approaches. However, careful examination of their results shows that
the poor performance is mainly isolated to the intercept terms. The
simulation reported in Table~3 of Hoffman, Sen and Weinberg (\citeyear{hoffsenwein2001}) does not
demonstrate bias for $b_1$, the covariate effect, and simulations
reported in Williamson, Datta and Satten (\citeyear{willdattsatt2003}) show bias on the order of $5\%$
(their Table 1) for covariate effects. Similar results hold for maximum
likelihood and random intercept models (Neu\-haus and McCulloch, \citeyear{neuhmccu2011}).

In summary, bias in estimates of between-cluster covariates has not
been demonstrated. Efficiency loss has been demonstrated, but only in
distributions quite far from non-normality, indicating a high degree of
robustness. The degree of robustness is similar to that of
normality-based tests of means.

\section{Estimation of an Intercept}

In many cases, linear mixed models assuming normality give unbiased or
nearly unbiased estimates of all the fixed effect parameters including
the intercept, even when the random intercepts distribution is
non-normal. However, the same is not true for nonlinear models. Both
theoretical and simulation studies have shown that estimates of the
intercept may be biased when the random effects distribution is far
from normal. Neuhaus, Hauck and Kalbfleisch (\citeyear{neuhhauckalb1992}) give a local approximation for
logistic models relating bias of estimators of the intercept under
assumed normality for the random intercepts to asymmetry of the random
effects distribution. Heagerty and Kurland (\citeyear{heagkurl2001}) show asymptotic relative bias
on the order of 20\% or more when the true distribution is gamma but
the assumed normal.\looseness=1

Similar results hold for informative cluster sizes.
Williamson, Datta and Satten (\citeyear{willdattsatt2003}) demonstrate bias of approximately $30\%$ in
estimating the intercept using independence generalized
estimating\break
equations using simulations for a logistic model. Similarly,
Neuhaus and McCulloch (\citeyear{neuhmccu2011}) demonstrate modest bias in estimating the intercept
using maximum likelihood methods. They give a theoretical argument as
to why bias is to be expected in the intercept but not other regression
parameters.

In summary, misspecification of the shape of the random effects
distribution can introduce a moderate to large bias in estimation of
the intercept. This will carry over to possible bias in estimation of
the mean value of the outcome for fixed covariate values. So, when
inference focuses on mean estimation or the intercept, care should be
taken with the specification of the random effects distribution.

\section{\texorpdfstring{Prediction of the Random Effect,~$\lowercase{b_i}$}{Prediction of the Random Effect, b_i}}

Much less work exists on the effect of misspecification on prediction
of the random effects. Magder and Zeger (\citeyear{magdzege1996}) demonstrated robustness of
predictions as gauged by mean square error of prediction in linear
mixed models and conclude, ``Differences in the performances of the
various methods in estimation of the random effects, $b$, were
generally not very large.'' In simulating the performance of\vadjust{\goodbreak} the
posterior mean estimates in logistic models, Agresti, Caffo and Ohman-Strickland (\citeyear{agrecaffohma2004}) did
not find a change in performance when the true distribution was varied
between uniform, exponential and normal and the assumed distribution
was normal. However, as noted above, this does not directly address the
question of misspecification. They did find a number of situations in
which assuming a~normal distribution suffered a moderate loss of
performance when the true distribution was a discrete two-point
distribution. Zhang et~al. (\citeyear{zhansongqugree2008}) (their Table~2) showed only
modest differences in the mean square error of prediction of random
slopes and intercepts when assuming normality and under a true log
gamma distribution for a linear mixed model.
Mc{C}ulloch and Neuhaus (\citeyear{mccuneuh2010}) investigated both linear mixed models and logistic
mixed models under a variety of assumed and true distributions, both by
theory and simulation, and found only modest impacts to
misspecification on the mean square error of prediction, using
posterior mean predictions for the linear mixed models and posterior
mode predictions for the logistic mixed models. Exceptions were when a
distribution with limited support was assumed, but the true
distribution had a wider range of support and for large random effects
variances and large cluster sizes.

The \textit{shape} of the distribution of the best predicted values is
a different matter. A number of authors
(Lesaffre and Molenberghs, \citeyear{lesamole1991}; Magder and Zeger, \citeyear{magdzege1996}; Zhang and Davidian,
\citeyear{zhandavi2001}) have demonstrated
convincingly it may not reflect the true underlying shape of the
distribution of the $b_i$, but instead the assumed distribution. We
also demonstrate this in our example. So, although the
\textit{performance} of the best predicted values (as gauged by the
mean square error of prediction) is robust, shapes of distributions of
best predicted values are not.

To recap, the shape of the distribution of the best predicted values is
highly sensitive to the assumed form of the distribution. This is
unfortunate because it means that diagnostics based on the empirical
distribution of the best predicted values (e.g., histograms or Q--Q
plots) are unreliable. However, the performance of either posterior
mean or posterior mode predicted values (as judged by overall mean
square error of prediction) are generally robust across a wide variety
of assumed distributions. Some loss of efficiency was evidenced with
large cluster sizes and large random effects variances, in which
definition of the correct random effects distribution is clearer.

\section{Estimate the Random Effects Variance}

Again, with regard to estimation of the random effects variance, there
is less research. Neuhaus, Hauck and Kalbfleisch (\citeyear{neuhhauckalb1992}) and
Heagerty and Kurland (\citeyear{heagkurl2001}) demonstrated little effect on estimation of the
random effects variance, with asymptotic relative bias on the order of
15\% or less. Agresti, Caffo and Ohman-Strickland (\citeyear{agrecaffohma2004}) demonstrated some situations,
mostly with larger random effects variances and large cluster sizes,
when there was efficiency loss with assuming a normal distribution,
when the true distribution was a discrete, two-point distribution.
Neuhaus and McCulloch (\citeyear{neuhmccu2011}) show via simulation in the informative cluster size
case that bias is slight.

As noted, this aspect of misspecification has not received the scrutiny
that the other aspects above have, perhaps because inference centering
on the random effects variance is somewhat less common. As a result,
the literature is far from definitive. Nevertheless, the basic
picture is similar to that of covariate effects: estimates of the
random effects variance appears relatively robust to misspecification
of the random intercept distributional shape.

\section{Misspecification of Other Aspects of~the Model}

We have mainly considered the situation in which only a single aspect
of the model, the shape of the random intercepts distribution, has been
misspecified. In practice, of course, all assumptions are violated to
at least a minor degree. What is the impact of simultaneous
misspecification of more than one aspect? There is little in the
literature to report, but previous work does inform on two aspects.
First, Neuhaus and McCulloch (\citeyear{neuhmccu2006}) show that use of conditional likelihood
methods and methods that separate covariates into between and within
components eliminate or reduce bias for the within-cluster covariate in
the situation where the random effects are associated with the
covariates. The argument works by showing that estimation of the
within-cluster covariate is essentially divorced from the random
effects distribution when using these methods, also making it
impervious to shape misspecification. Second, the arguments of
Neuhaus and McCulloch (\citeyear{neuhmccu2011}), showing that informative cluster sizes have little
impact on the bias of regression coefficients, proceed by converting
the informative cluster size problem into the misspecified random
effects distribution as displayed in (\ref{eqtruecondit1}). This same
argument can be used to show that, even with simultaneous
misspecification of the shape of the random effects distribution and
informative cluster sizes, maximum likelihood estimators of the
regression coefficients (other than the intercept) will be consistent
in a~li\-near mixed model and, for $\beta=0$, consistent at zero for
generalized linear mixed models. While not all-encompassing, this
result for generalized linear mixed models is important for testing the
null hypothesis that $\beta=0$.

\section{A Simulation Study}

We performed a simulation study to evaluate the performance of each of
the above aspects of inference. The goal was to gauge the performance
of assumed normal fits to a distribution that was highly non-normal,
but not as extreme as a two-point, discrete distribution. We chose a
$\operatorname{Tukey}(g,h)$ distribution, since, depending on the values of $g$ and
$h,$ the Tukey distribution can be quite skewed and/or heavy-tailed.
See He and Raghunathan (\citeyear{heragh2006}) for a recent reference. We chose $g=0.5$ and
$h=0.1$, which gives a mean of 0.31, variance of 2.27, skewness of
3.41 and a kurtosis of 44.24.

The simulations used two covariates: one within-cluster and one
between-cluster covariate. The within-cluster covariate was equally
spaced between 0 and~1. The between-cluster covariate was binary with
a~25\%/\break 75\% division. The parameter values were: $\beta_0=-2.5$,
$\beta_{\mathrm{between}}=2$, $\beta_{\mathrm{within}}=1$, $\sigma_b=1$. We set the number of
clusters, $m,$ to 200 and used a variety of cluster sizes ($n=2, 4, 6,
10, 20$ and $40$). We simulated data under a~logistic link and the
$\operatorname{Tukey}(g,h)$ distribution. We ran 1,000 replications for each
scenario and used common random numbers across different cluster sizes
and fitted distributions to increase precision of comparisons. We
conducted simulations in SAS (Ver~9.1, SAS Institute, Cary NC) and fit
models using Proc NLMIXED.

\begin{figure*}

\includegraphics{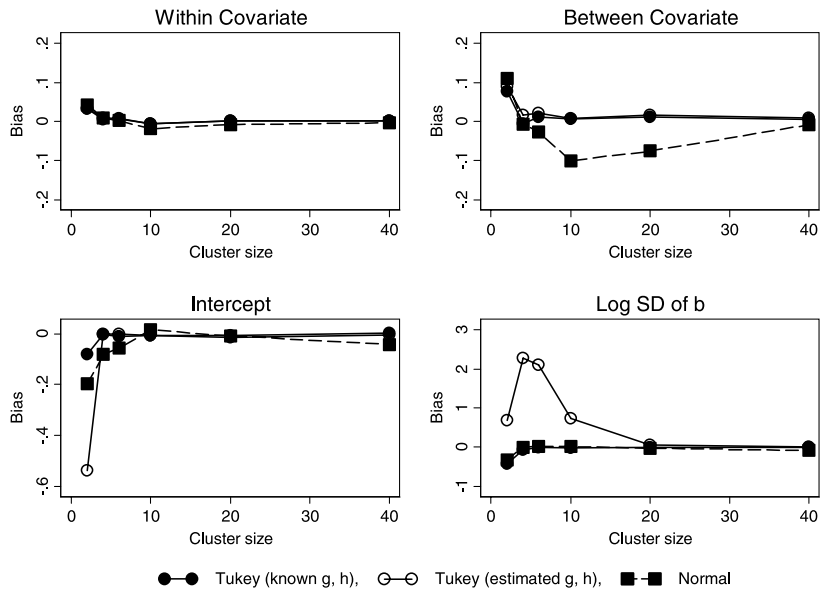}

\caption{Bias of estimators of the parameters of a logistic mixed
effects model, (\protect\ref{glmm}), with $\operatorname{Tukey}(g,h)$ distributed random
intercepts, fit with assumed normal and Tukey distributions.}\label{figbias}
\end{figure*}

To each simulated data set we fit three GLMMs with a logistic link.
The first model assumed that the random effects were standard normal
and the second assumed they followed a standardized $\operatorname{Tukey}(g,h)$
distribution with unknown $g$ and $h.$ Especially with small cluster
sizes, there is little or no information about the shape parameters of
the Tukey distribution and trying to estimate those parameters led to
unstable estimation. To directly assess the effect of misspecification
and to avoid confounding the effects of an incorrect distributional
shape with the estimation of the two additional parameters, we fit a
third GLMM with the random effects following a Tukey distribution with
$g$ and $h$ fixed at their true values of 0.5 and 0.1, but still
estimating the random effects variance.

Figure \ref{figbias} gives the results for the bias of the estimators.
As expected, there was virtually no impact of using an assumed normal
distribution in the estimation of the within-cluster covariate. There
was a~modest impact in estimating the between covariate and the
intercept (but with bias less than $5\%$ for the between-cluster
covariate and less than $10\%$ for the intercept). Impact on
estimation of the log standard deviation of the random effects was
negligible.

As noted, estimation of the $g$ and $h$ parameters for the Tukey
distribution introduced instability in estimating the log standard
deviation of the random effects. This led to poor estimation of the
intercept for cluster size $n=2$ and for the log standard deviation for
cluster sizes $n=2, 4, 6$ and $10$. Perhaps surprisingly, the
estimation of $g$ and $h$ had little impact on estimation of
$\beta_{\mathrm{between}}$.

Fitting these models is quite involved, so we give some further details
on the simulation. Not all the replications in the simulations
achieved convergence via NLMIXED. Fitting the assumed normal and Tukey
distribution with known $g$ and $h$ had excellent convergence rates,
all above $95\%.$ However, the Tukey distribution with estimated $g$
and $h$ was numerically challenging and convergence rates for $n=4$ and
$6$ were lower ($82.5\%$ and $83.8\%,$ respectively), with estimation
of $\log\sigma_b$ especially problematic. The figures give results
for all the runs, using final parameter values in cases with
nonconvergence. Comparing the convergent runs only with all the runs
showed little impact on the bias of the parameters in Figure
\ref{figbias}, excepting $\log\sigma_b.$
Simulation code is available from the authors.

\begin{figure*}

\includegraphics{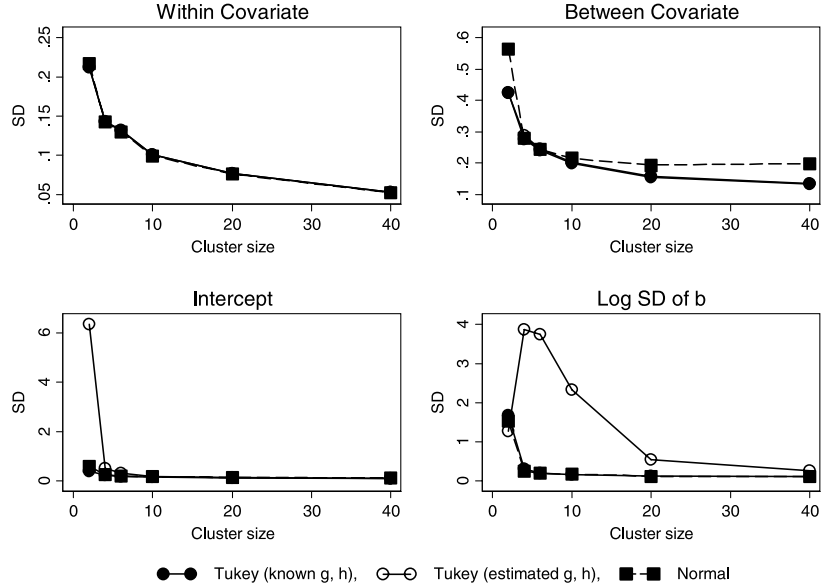}

\caption{Standard deviation of estimators of the parameters of a
logistic mixed effects model, (\protect\ref{glmm}), with $\operatorname{Tukey}(g,h)$
distributed random intercepts, fit with assumed normal and Tukey
distributions.}\label{figsd}
\end{figure*}

Figure \ref{figsd} gives the results for the standard deviations of the
estimators.
The standard deviations of the estimators were nearly the same under
both the fitted normal and Tukey distributions across all the cluster
sizes and for all the parameters, with the exception being a modest
loss of efficiency for the between-cluster covariate, especially with
larger cluster sizes. So misspecifying the random effects distribution
produced essentially no loss in estimation efficiency.

\begin{figure}

\includegraphics{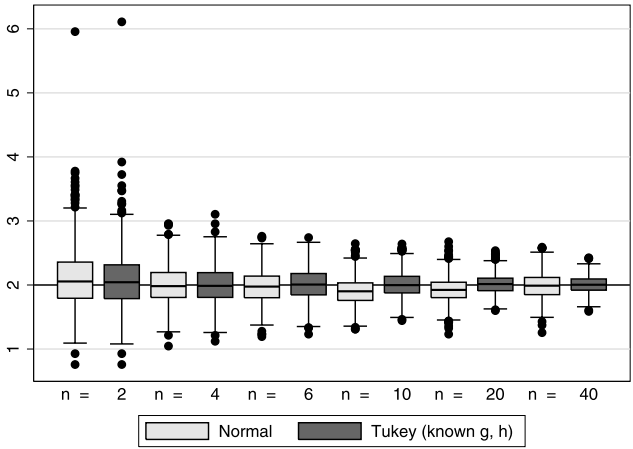}

\caption{Boxplots of estimates of $\beta_{\mathrm{between}}$ from a logistic
mixed effects model, (\protect\ref{glmm}), with $\operatorname{Tukey}(g,h)$ distributed random
intercepts, fit with assumed normal and Tukey distributions. Reference
line given at the true value of $\beta_{\mathrm{between}}=2$.}\label{figbox}
\end{figure}

Since estimation of between-cluster covariates is an important
inferential goal in clustered data settings and there is a modest
impact of misspecifying the distribution as normal, we give more detail
on the actual distributions of the estimates of the between-cluster
covariate effect. Figure \ref{figbox} shows boxplots of the estimates
under the assumed normal and Tukey (with fixed $g$ and $h$) fits.
Behavior of the assumed normal fit is slightly worse than the true
Tukey fit for cluster sizes of $10$ and $20$ and somewhat more variable
for a cluster size of $40.$ For smaller cluster sizes the results are
comparable. Given the extreme differences between a normal and
$\operatorname{Tukey}(0.5,0.1)$ distribution, this represents a large degree of
robustness, especially for cluster sizes 6 or smaller.

%

%
\begin{figure}
\vspace*{14pt}
\includegraphics{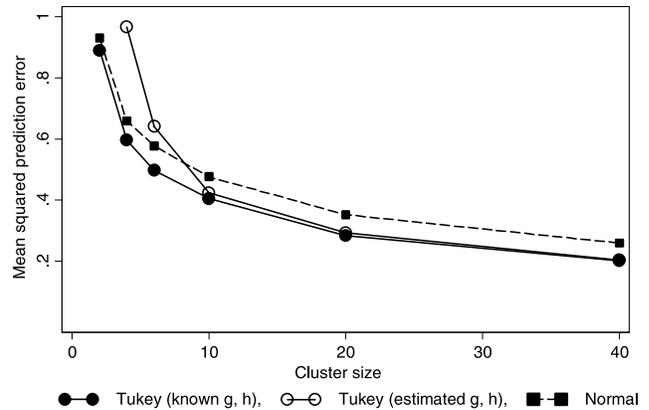}

\caption{Mean square error of prediction of the random effects from a
logistic mixed effects model, (\protect\ref{glmm}), with $\operatorname{Tukey}(g,h)$
distributed random intercepts, fit with assumed normal and Tukey
distributions.}\label{figmsep}
\end{figure}

SAS Proc NLMIXED calculated best predicted values as modes of the
$\log$ of $f(\mathbf{y}_i | \mathbf{x}_i, b_i)f_b(b_i | \sigma_b)$. Figure
\ref{figmsep} shows the mean squared error of prediction under the
three fitted models. Compared to the Tukey distribution with estimated
$g$ and $h$, the misspecified normal actually outperforms it up until
cluster sizes of $10.$ Compared to the Tukey distribution with fixed
$g$ and $h$, the performance is slightly worse, with the inflation in
mean square error of prediction ranging from about $5\%$ at $n=2$ to
about $20\%$ for the larger cluster sizes. Again, given the extreme
differences between a normal and Tukey, this is a high degree of
robustness.

\begin{table*}
\tabcolsep=0pt
\tablewidth=350pt
\caption{HERS model fit
comparisons with different assumed random effect distributions.
Model-based standard errors for the fixed effects are given as subscripts}\label{tabhersmodelsbp}
\begin{tabular*}{350pt}{@{\extracolsep{4in minus 4in}}lc@{\hspace*{-1pt}}ccccc@{}}
\hline
\multirow{2}{80pt}[-8pt]{\textbf{Random effects distribution}} && \multicolumn{5}{c@{}}{\textbf{Parameter
estimates}}\\
\ccline{3-7}\\ [-7pt]
 & $\bolds{-2}$ \textbf{loglik} & \textbf{Intercept} & \textbf{Visit}& \textbf{BMI} & \textbf{HTN} & $\bolds{\log(\hat
{\sigma}_b)}$ \\
\hline
Normal & 3695 & $-4.28_{0.41}$& $0.86_{0.05}$& $0.024_{0.013}$&
$-0.36_{0.16}$& 0.49\\
Exponential & 3732 & $-3.96_{0.38}$ &$0.84_{0.05}$& $0.023_{0.013}$&
$-0.30_{0.14}$& 0.19 \\
Discrete & 3674 & $-4.05_{0.38}$ &$0.87_{0.05}$& $0.021_{0.012}$&
$-0.37_{0.16}$ &0.23\\
Tukey & 3677 & $-4.10_{1.06}$ &$0.86_{0.05}$& $0.022_{0.013}$&
$-0.36_{0.16}$ &0.28 \\
\hline
\end{tabular*}
\end{table*}

These results support the broad conclusion that, for distributional
shapes quite different from the normal, estimation of the intercept may
be biased. However, the bias for estimating other parameters was low
and efficiency high. For prediction of random effects, the mean square
error of prediction was modestly increased when incorrectly assuming
normality, especially for larger cluster sizes.

\section{Example---HERS}\label{sechers}
HERS was a randomized, blinded, placebo control\-led trial for women with
previous coronary disease. The study enrolled 2,763 women and followed
them annually for five subsequent visits. We will consider only the
subset of $N=1{,}378$ that were not diabetic and who had systolic blood
pressure less than 140 at the beginning of the study. We treat HERS as
a prospective cohort study using the first four visits. We modeled the
binary outcome of high blood pressure (HBP), defined as a systolic
blood pressure greater than 140, as a function of visit (numerical 0
through~3), body mass index (BMI) and whether or not the participant
was on high blood pressure medication at that visit (HTN). The visit
trend variable is essentially a~within-person covariate and BMI and HTN
are mostly between-person covariates (with $95\%$ and $76\%,$
respectively, of their variability associated with between-person
differences).

With $\mbox{HBP}_{it}$ equal to $1$ if woman $i$ at visit $t$ had high
blood pressure (and $0$ otherwise), our random intercepts logistic
regression model was
%
\begin{eqnarray}\label{hersbpmodel}
&&\mbox{logit} (\mbox{P}\{\mbox{HBP}_{it}=1\} )\nonumber\\
&&\quad= \beta
_0 + b_{0i} + \beta_1 \mbox{visit}_t + \beta_2 \mbox{BMI}_{it} +
\beta_3\mbox{HTN}_{it}, \nonumber\\ [-8pt]\\ [-8pt]
\eqntext{\mbox{where }   b_{0i} \sim \mbox{i.i.d. } \mathcal{N}(0, \sigma
^2_b)\quad \mbox{or}} \\
\eqntext{b_{0i} \sim \mbox{i.i.d. }\sigma_b\{\mbox{Exp}(1) - 1\}\quad \mbox{or}} \\
\eqntext{b_{0i} \sim \mbox{i.i.d. } \mbox{discrete with three mass
points}\quad \mbox{or}}\\
\eqntext{b_{0i} \sim \mbox{i.i.d.
}\sigma_b\{\operatorname{Tukey}(g,h)\}.}
\end{eqnarray}

We fit models via maximum likelihood to the data using, in turn, each
of the assumed random effects distributions given in
(\ref{hersbpmodel}). We used SAS Proc\break NLMIXED (SAS Institute, Cary,
NC) for the continuous distributions, which gives posterior mode
estimates of the $b_{0i}.$ We used the Stata module\break GLLAMM
(\href{http://www.gllamm.org}{www.gllamm. org}) to fit the three point discrete discrete distribution,
which gives posterior mean estimates of the $b_{0i}.$

We chose the exponential ($\mbox{Exp}$) and Tukey distributions since
they are parametric but quite different from the normal. The
exponential distribution has scale parameter 1, a bounded range, and is
skewed right. The Tukey distribution is a~flexible parametric family
that can take on a wide variety of shapes (He and Raghunathan, \citeyear{heragh2006}); it
includes the normal as a special case ($g=h=0$) but also includes
distributions with extreme skewness and kurtosis. The discrete
distribution is similar to a nonparametric maximum likelihood fit, but
specifies a priori the number of mass points.

Table \ref{tabhersmodelsbp} lists the fitted coefficients and maximized
log likelihood values.
As expected, the fixed effects parameter estimates are quite similar,
especially the ``within'' covariate, Visit, even though there are
modest differences in the fits of the models as judged by the value of
the maximized log likelihood. The shape parameters for the Tukey
distribution had large standard errors but were estimated to be\break $g=2.5$
and $h=-1.8$, which is a~distribution with boun\-ded range, and is
slightly skewed right and hea\-vy tailed (skewness of about 0.9 and
kurtosis of about~2.3).

\begin{figure*}

\includegraphics{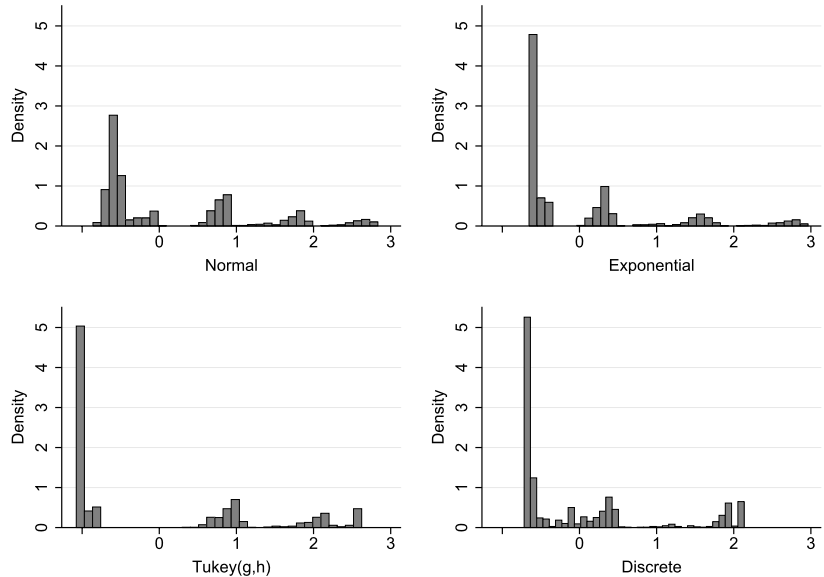}

\caption{Best predicted values for the HERS data under different
assumed random effects distributions.}\label{figbpplots}
\end{figure*}

Figure \ref{figbpplots} gives histograms of the best predicted values
of the random intercept deviations, $b_{0i},$ under each of the above
assumed distributions.
As is evident from the figure, the shape of the distribution of the
best predicted values is dependent on the assumed random effects
distribution, with the discrete and exponential distributions
especially having different shapes than the normal and Tukey, which are
similar to one another.

%
%

\section{Random Intercepts and Slopes}\label{secrslopes}

The remarks above pertain mostly to models with random intercepts only.
An understudied area is what happens with the further complication of
random coefficient models, such as random intercepts and slopes. The
line of argument cited above using minimization of Kullback--Leibler
divergence (Neuhaus, Hauck and Kalbfleisch, \citeyear{neuhhauckalb1992}; Neuhaus, Kalbfleisch and Hauck,
\citeyear{neuhkalbhauc1994}; Heagerty and Kurland, \citeyear{heagkurl2001}) can
also be used in the random intercepts and slopes situation. This line
of argument shows, for the linear mixed model, that estimates of fixed
effects parameters will be consistent. However, for the linear mixed
model this is well known (Verbecke and Lesaffre, \citeyear{verblesa1997}) by alternate means. For
generalized linear mixed models, results analogous to the random
intercepts case hold. For fixed effects for which there is a
corresponding, misspecified random effects distribution, bias can
result. But fixed effects orthogonal to random effects are little
affected. For example, if a covariate is both a fixed effect and a
random effect (i.e., a random slope), the estimate of the fixed effect
for that covariate may exhibit bias when the distribution of the random
slope is misspecified. But estimates of covariate effects, orthogonal
to the one which is random, are little affected. For the informative
cluster size situation, Neuhaus and McCulloch (\citeyear{neuhmccu2011}) make these arguments more
precise. In the case of prediction of realized values or random
effects, Mc{C}ulloch and Neuhaus  (\citeyear{mccuneuh2010}) report limited results for random
intercepts and slopes; they show a fair degree of robustness for mean
square error of prediction and comparable behavior to the random
intercepts case.

Contrary to the above, Liti{\`{e}}re, Alonso and Molenberghs (\citeyear{litialonmole2008}) describe some
simulations for the random intercepts and slopes situation that report
to show significant bias for a mixed effects logistic mo\-del with true
normal or true mixture of normal distributions to which an assumed
bivariate normal distribution is fit. The situation they simulate is
quite extreme. The true bivariate mixture distribution corresponding
to their least extreme configuration, which they call $\mathbf{V}_1$, is a~mixture of standard normals with
mean values at plus or minus 2 for
both the random intercept and slope; this generates a distribution (in
two dimensions) with two isolated peaks. The standard deviation on the
logit scale for both the intercepts and slopes for this model is
$\sqrt{5} \approx 2.24.$

%
\begin{table*}
\tabcolsep=0pt
\caption{Medians of maximum likelihood estimates of fixed effects from fitting
model (\protect\ref{litmodel}) assuming a bivariate normal distribution when
the true distribution is normal or a mixture of bivariate normals.
$\beta_0$ is the intercept, $\beta_b$ is the between-cluster
coefficient and $\beta_w$ is the within-cluster coefficient. In each
case the random intercepts had variance 5 and a correlation with the
random slopes of 0.9. The random slopes variance was either 5 (upper
panel) or the more reasonable value of 0.08 (lower panel). The
simulation generated 500 replications with 100 clusters of size 6}\label{tabsimresults}
\begin{tabular*}{\textwidth}{@{\extracolsep{4in minus 4in}}lcccccc@{}}
\hline
& & &\multicolumn{3}{c}{\textbf{Median estimates (true values)}}\\
\ccline{4-6}\\ [-7pt]
& \textbf{True dist'n}&\textbf{Simulation} &  $\bolds{\beta_0}$ \textbf{($\bolds{-}$6)}& $\bolds{\beta_b}$
\textbf{(2)}& $\bolds{\beta_w}$ \textbf{(1)} &
\textbf{Convergence rate}\\
\ccline{2-7}\\[-7pt]
$\operatorname{var}(b_{wi})=5$ & Normal & Current& \phantom{0}$-$6.26& 2.13& \phantom{$-$}1.07& 98\%\\
& Normal & Litiere et al.& \phantom{0}$-$6.14& 2.04& \phantom{$-$}1.04& not given\\
& Mixture & Current& \phantom{0}$-$6.34& 2.87& \phantom{$-$}0.86& 99\%\\
& Mixture & Litiere et al.& $-$10.52&2.52&$-$0.13& ${<}43$\%\ \\ [5pt]
$\operatorname{var}(b_{wi})=0.08$ & Normal & Current &\phantom{0}$-$6.18& 2.12& \phantom{$-$}1.04& 100\%\\
& Mixture & Current &\phantom{0}$-$5.89& 2.26 &\phantom{$-$}0.95&100\%\\
\hline
\end{tabular*}
\vspace*{-5pt}
\end{table*}

For the random intercept, a random effect one standard deviation above
its mean of 0 would have an odds of the outcome that is $\exp(2.24)
\approx 9.36$ higher than a~random effect at its mean of 0. So this is
a moderately large effect. The effect of a random slope one standard
deviation above its mean of 0 depends on the values of the associated
covariate, which range from 0 to 8 in their simulation. So a random
slope one standard deviation above its mean of 0 would have the same
odds of the outcome at $t=0$ compared to an\vadjust{\goodbreak} ``average'' random slope,
but at $t=8$ would have an odds of the outcome of
$\exp(8\sqrt{5})\approx58{,}700{,}000$ times higher. Furthermore, the
correlation between the random intercepts and slopes was 0.9. The other
scenarios they simulate are even more extreme! A more reasonable value
for the random slope variance would be approximately 0.08. At $t=8$
this would generate an odds\vspace*{1pt} of an outcome which is
$\exp(8\sqrt{0.08})\approx9.61$ times higher, an effect comparable in
magnitude to the random intercepts.

They also report poor convergence rates when fitting the assumed
bivariate normal distribution to the bivariate mixture distribution. We
performed a~small simulation to check their results and to assess the
magnitude of the bias under the highly non-normal but more reasonable
situation where the random slopes variance was 0.08. We used the same
model as Liti{\`{e}}re, Alonso and Molenberghs (\citeyear{litialonmole2008}), namely,
\begin{eqnarray}\label{litmodel}
&&\mbox{logit} (P\{Y_{it}=1\} )\nonumber\\ [-8pt]\\ [-8pt]
&&\quad = (\beta_0+ b_{0i}) +
\beta_bz_i+(\beta_w + b_{wi})t_j,\nonumber
\end{eqnarray}
with $z_i$ binary with equal
proportions of 0 and 1, $t_j=0, 1, 2, 4, 6$, and 8 (within each cluster
of size 6) and 100 clusters. We performed 500 replications and fit the
models via maximum likelihood using SAS Proc NLMIXED assuming a
bivariate normal distribution. Table \ref{tabsimresults} reports the
results.
When fitting to a true normal distribution, the median values when the
random slope variance is 5 (upper panel) are remarkably close to the
true values. When fitting to a~true mixture distribution, the median
values exhibit bias, but not nearly as extreme as those reported by
Liti{\`{e}}re, Alonso and Molenberghs (\citeyear{litialonmole2008}). Convergence rates when fitting to
the mixture
distribution are much better as well.\vadjust{\goodbreak} Our results are consistent with
previous literature, in that the impact on the within-cluster
coefficient $\beta_w$ is less extreme than the between-cluster
coefficient, $\beta_b$. Under the more reasonable random slopes
variance of 0.08 there is only a small amount of bias. Of course, it is
important to remember that, even under the more reasonable value of the
variance, the mixture model is a highly non-normal bivariate
distribution with most of its mass concentrated around ($-$2, $-$2) and (2, 2).

\section{Summary}
We considered robustness to the assumed distribution for a random
intercept when using maximum likelihood methods for fitting a
generalized linear mixed model such as (\ref{glmm}). In practice, data
analysts commonly assume normality for the random effects. Theory and
simulation studies indicate that most aspects of statistical inference
are highly robust to this assumption. Especially robust were
inferences for within-cluster covariate effects, which are often a key
reason for considering a clustered data design.

These conclusions are contrary to much of the previous literature. We
have argued that this is because (a)~results for the nonclustered data
situation (e.g., Heckman and Singer, \citeyear{hecksing1984}) are incorrectly interpreted as
relevant to the clustered data setting, (b) results restricted to one
portion of the model (primarily the intercept) are characterized as a
general failure of the methodology, and (c) reanalysis of previous
scenarios has sometimes given different interpretations or results.

Exceptions to this robustness include estimation of the intercept,
which can be biased when the random intercept distribution is
misspecified.\vadjust{\goodbreak} The shape of the estimated random effects distribution
(based on the distributional shape of the predicted random effects) can
also be quite sensitive to the shape of the assumed distribution and
thus not reflect the true distribution.

However, a wide array of inferences are quite robust to this type of
misspecification, including estimation of covariate effects, estimation
of the random effects variance and the mean square error of prediction
of the realized value of random effects. We also argued that this
robustness extends to random intercept situations where the cluster
size is informative.

Between-cluster covariate effects and estimation of the log of the
standard deviation of the random effects were found to be robust, but
not impervious, to misspecification of the random effects distribution.
The literature and results here indicate that a~modest amount of bias
and minor loss of efficiency can occur, especially when the true
distribution is far from the assumed (e.g., assuming a normal
distribution when the true distribution is a two-point distribution or
the Tukey distribution investigated here), the random effects are
large, and the cluster size is large. When random effects and cluster
sizes are large, the data provide a fair amount of information as to
the shape of the random effects distribution and correctly specifying
the random effects distribution gives some advantage.

Particularly robust was estimation of within-clus\-ter covariate effects.
This is especially important because exploitation of within-cluster
comparisons is often the rationale for conducting longitudinal or
clustered data studies. Virtually all the evidence in the literature
and the new results reported here indicate that within-cluster
covariate effects are estimated with no more bias when fitting a
misspecified model than when fitting the correct model and with little
or no loss of efficiency.

\section*{Acknowledgments}
We thank Ross Boylan for computational assistance with the
simulation studies and Stephen Hulley for use of the HERS
data set.  Support was provided by NIH Grant R01 CA82370.


\end{document}